\documentclass[aps,pre,twocolumn,groupedaddress,superscriptaddress,showpacs]{revtex4-1}

\usepackage{amsmath}
\usepackage{amssymb}
\usepackage{graphicx}
\usepackage{float}
\usepackage{color}

\begin{document}

\title{Relaxation dynamics of functionalized colloids on attractive substrates}

  \author{C. S. Dias}
   \email{csdias@fc.ul.pt}
    \affiliation{Centro de F\'{\i}sica Te\'orica e Computacional, Departamento de F\'{\i}sica, Faculdade de Ci\^{e}ncias, 
    Universidade de Lisboa, Lisboa, Portugal}

  \author{C. Braga}
   \email{c.correia-braga@imperial.ac.uk}
    \affiliation{Department of Chemical Engineering, Imperial College London, London SW7 2AZ, United Kingdom}    
    
  \author{N. A. M. Ara\'ujo}
   \email{nmaraujo@fc.ul.pt}
   \affiliation{Centro de F\'{\i}sica Te\'orica e Computacional, Departamento de F\'{\i}sica, Faculdade de Ci\^{e}ncias, 
   Universidade de Lisboa, Lisboa, Portugal}

  \author{M. M. Telo da Gama}
   \email{mmgama@fc.ul.pt}
    \affiliation{Centro de F\'{\i}sica Te\'orica e Computacional, Departamento de F\'{\i}sica, Faculdade de Ci\^{e}ncias, 
    Universidade de Lisboa, Lisboa, Portugal}

\begin{abstract}
Particle-based simulations are performed to study the post-relaxation dynamics of 
functionalized (patchy) colloids adsorbed on an attractive substrate. Kinetically 
arrested structures that depend on the number of adsorbed particles and the
strength of the particle-particle and particle-substrate interactions are identified. The radial 
distribution function is characterized by a sequence of peaks, with relative intensities that depend
on the number of adsorbed particles. The first-layer coverage is a 
non-monotonic function of the number of particles, with an optimal value around one 
layer of adsorbed particles. The initial relaxation 
towards these structures is characterized by a fast (exponential) and a slow (power-law) 
dynamics. The fast relaxation timescale is a linearly increasing function of the 
number of adsorbed particles in the submonolayer regime, but it saturates for more 
than one adsorbed layer. The slow dynamics exhibits two characteristic exponents,
depending on the surface coverage.
\end{abstract}

\maketitle

\section{Introduction}

Large-scale production of materials with enhanced physical properties from the
self-organization of colloidal particles is believed to set the stage for a 
revolution in materials engineering
\cite{Zhang2014a,Paulson2015,Lu2013,Duguet2011,Hu2012,Kretzschmar2011,Sacanna2011,Solomon2011,Pawar2010,Bianchi2011,Sacanna2013a,Manoharan2015}. 
There has been a sustained effort towards finding 
design rules to obtain the desired structures through the control of the experimental 
conditions and the particle-particle interactions. One promising route is to promote the 
aggregation of anisotropic colloidal particles on an attractive substrate \cite{Iwashita2014,Iwashita2013,VanOostrum2015}. However, the 
feasibility of the target structures is deeply compromised by the kinetic pathways of 
aggregation and post-relaxation. In this work, we study the kinetics
of relaxation of patchy particles on a flat substrate.

Patchy particles usually refer to spherical colloids with a chemical decoration (patches) on their surface. The idea is to control 
the directionality and strength of the particle-particle interaction, the maximum valence of the colloid, and the local structure of the aggregates. 
The functionalization of the patches can be obtained using, for example, DNA, polymers, or enzymes 
\cite{James2015,Peng2014,Wilner2012,Rogers2015,Geerts2010}. This type of functionalization allows
a very detailed control on of colloidal valence, strength, and selectivity of the colloidal bonds. 
During self-assembly, the chemical nature (or strength) of the patch-patch 
interaction leads to large energy barriers impeding the formation of equilibrium structures, 
promoting kinetically trapped structures, such as gels, glasses or polycrystals \cite{Bianchi2013,Khan2015,Guo2014,Markova2014,Vasilyev2013,Zhang2014a}. 
One route to control the assembly of non-equilibrium structures is the use of flat or patterned substrates \cite{Araujo2008,Cadilhe2007,Eskandari2014,Silvestre2014}. 
Recent studies of self-assembly of patchy particles 
on substrates have focused on the thermodynamic structures \cite{Bernardino2012, Kalyuzhnyi2015,Sokolowski2014,Pizio2014,Rzysko2015}
or on the fully irreversible regime \cite{Dias2013a,Dias2014,Dias2014a,Araujo2015}. 
Here, we study the relaxation dynamics on the substrate and how it depends on the number of adsorbed particles and the strength of 
the particle-substrate interaction. This article is organized in the following way. In the next section, the model and the 
simulation parameters are described. Results are discussed in Sec.~\ref{sec.res} and in Sec.~\ref{sec.conc} we draw some conclusions.

\section{Model and Numerical Simulations}\label{sec.model}

We consider a three dimensional system of spherical colloidal particles with three patches 
distributed along the equator, with an opening angle of $2\pi/3$. Colloidal particles are represented by 
a core of mass $m_c$ and three patches at a distance $R$ from the center. 
The particles representing the patches are of zero diameter and mass $10^{-5}m_c$. 
Their relative position to the center of the core is fixed at all times. 

  \begin{figure}[t]
   \begin{center}
    \includegraphics[width=\columnwidth]{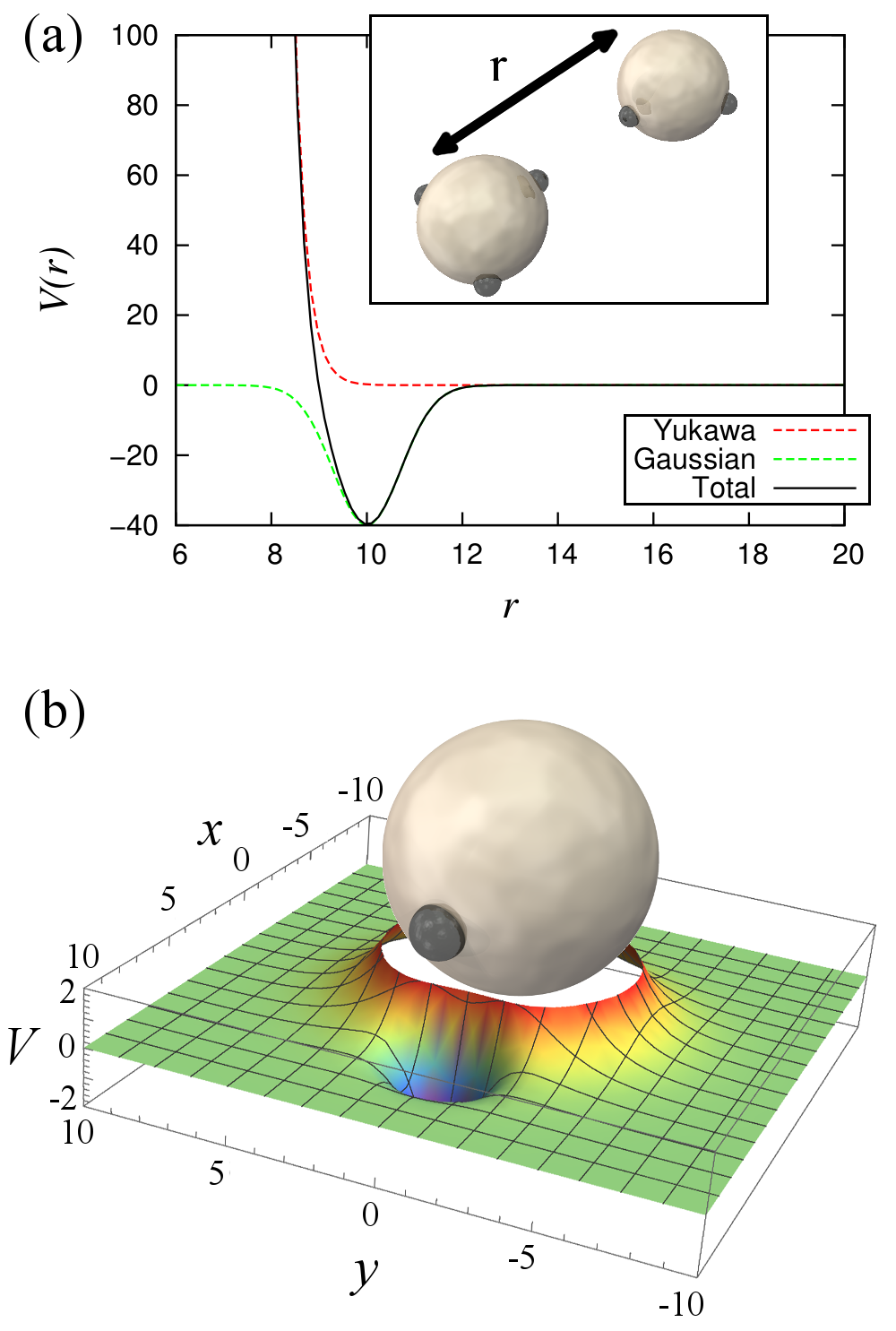} \\
\caption{(a) Schematic representation of the core-core repulsive interaction (Yukawa) and the patch-patch attraction (Gaussian) as a 
function of the distance between the center of two patchy particles when their patches are aligned. (b) Energy landscape of a 
single-patch particle interacting with one single patch as a probe at a position (x,y) relative to the center of 
the patchy particle.}
  \label{fig.potentials}
   \end{center}
  \end{figure} 

The core-core interaction is repulsive, described by a
Yukawa potential (see Fig.~\ref{fig.potentials}(a)),
   \begin{equation}
    V_Y(r)=\frac{A}{k}\exp{\left(-k\left[r-(R_i+R_j)\right]\right)}, \label{eq.Yukawa}
   \end{equation}
where $R_i$ and $R_j$ are the effective radii of the two interacting particles, $A=1$ is the
interaction strength and $k=4$ the inverse of the screening length. The core-core interaction 
is truncated at a cutoff distance $r_c=3R$ (at $r_c$ the potential is $10^{-9}A/k$). 
The patch-patch interaction is described by an attractive Gaussian potential \cite{Vasilyev2013}, 
(see Fig.~\ref{fig.potentials}(a)),
   \begin{equation}
    V_G(r_p)=-\epsilon\exp(-\sigma r_p^2),
   \end{equation}
where $\epsilon=40$ (in units of $k_BT$) is the interaction strength, $\sigma=0.1R$ the width of the Gaussian, 
and $r_p$ is the distance between the two interacting patches. The patch-patch interaction is truncated at a cutoff distance $r_{pc}=2R$.
The superposition of these two interactions and the resulting energy landscape are represented in Figs.~\ref{fig.potentials}(a)~and~(b).
For these parameters, we expect at most one bond per patch. However, as we discuss below, 
in certain cases we observe the formation of double bonds, i.e. two bonds per patch.

 \begin{figure*}[t]
   \begin{center}
    \includegraphics[width=1.8\columnwidth]{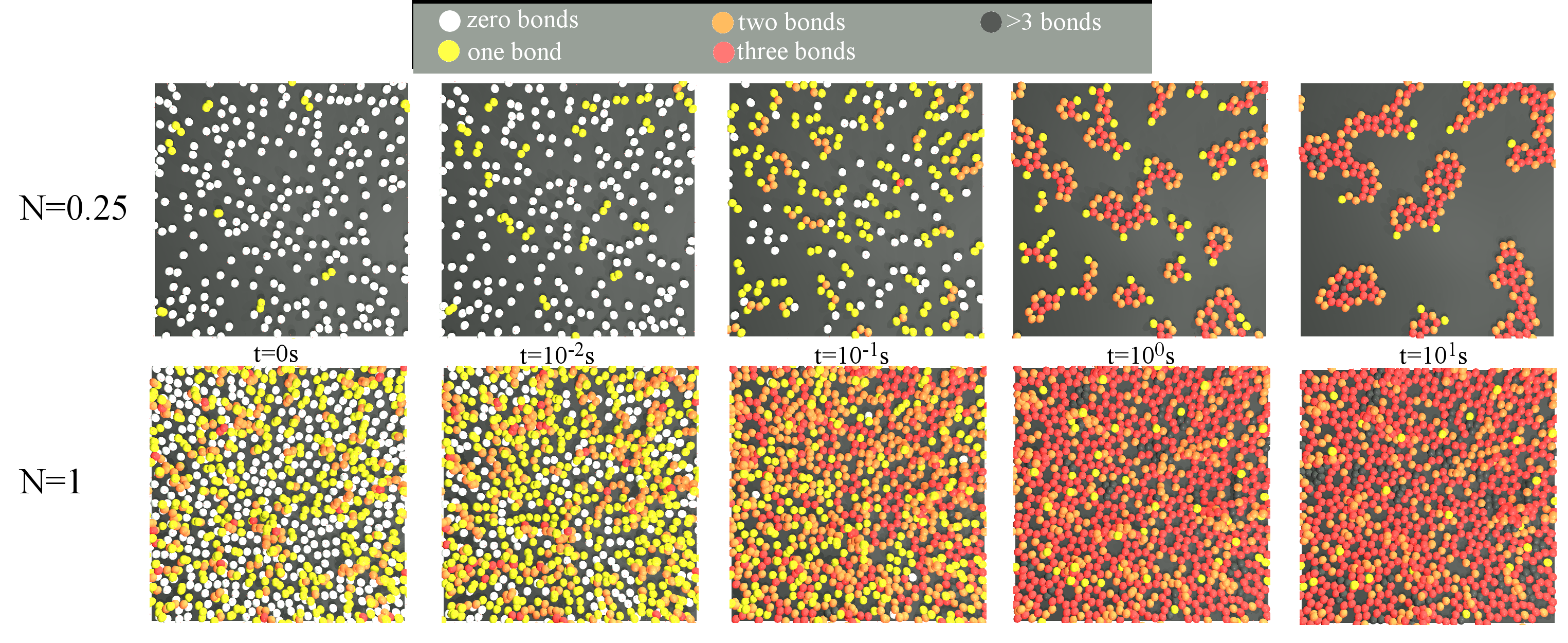} \\
\caption{Snapshots for $N=0.25$ and $N=1$. 
Images generated at $t=\{0, 10^{-2}, 10^{-1}, 10^0, 10^1\}$ seconds
on a substrate of lateral size L=32. Colors indicate the number of bonds of the colloidal particle.}
  \label{fig.substrate_snaps_con}
   \end{center}
  \end{figure*}

We consider an attractive substrate interacting isotropically with the patchy particles. The 
pairwise potential was derived from the Hamaker theory for two spheres \cite{Everaers2003} in the 
limit where the radius of one particle diverges. This gives,
\begin{equation}
 V_A=-\frac{A_H}{6}\left[\frac{2R(R+D)}{D(D+2R)}+\ln\left(\frac{D}{D+2R}\right)\right], \label{eq.Hamaker_Att}
\end{equation}
for the attractive part and
\begin{equation}
 V_R=\frac{A_H\sigma^6}{7560}\left(\frac{6R-D}{D^7}+\frac{D+8R}{(D+2R)^7}\right), \label{eq.Hamaker_Rep}
\end{equation}
for the repulsive one, where $A_H$ is the Hamaker's constant and $D=r-R$ the distance between the 
surface of the particle and the substrate. The potential is highly repulsive 
for distances shorter than the particle radius and is attractive at longer distances. It has a minimum at $r\approx1.1R$.

To resolve the stochastic trajectories of the particles, we describe them as rigid bodies and integrate the corresponding Langevin 
equations of motion for the translational degrees of freedom,
\begin{equation}
 m\dot{\vec{v}}(t)=-\nabla_{\vec{r}} V(\vec{r})-\frac{m}{\tau_t}\vec{v}(t)+\sqrt{\frac{2mk_BT}{\tau_t}}\vec{\xi}(t), \label{eq.trans_Langevin_dynamics}
\end{equation}
and rotational ones,
\begin{equation}
 I\dot{\vec{\omega}}(t)=-\nabla_{\vec{\theta}} V(\vec{\theta})-\frac{I}{\tau_r}\vec{\omega}(t)+\sqrt{\frac{2Ik_BT}{\tau_r}}\vec{\xi}(t).\label{eq.rot_Langevin_dynamics}
\end{equation}
The equations are integrated using the velocity
Verlet scheme with a time step of $\Delta t=0.01$  and Large-scale Atomic/Molecular Massively Parallel Simulator (LAMMPS) 
for efficient parallel simulations \cite{Plimpton1995}.
$\vec{v}$ and $\vec{\omega}$ 
are the translational and angular velocity, $m$ and $I$ are
mass and inertia of the patchy particle, $V$ is the pairwise potential, and $\vec{\xi}(t)$ is the stochastic term, 
from the thermal noise, given from a random distribution of zero mean.
We consider the damping time for the translational motion,
\begin{equation}
 \tau_t=\frac{m}{6\pi\eta R}, \label{eq.damping}
\end{equation}
which we set to be $\tau_t=0.1\Delta t$. From the Stokes-Einstein-Debye
relation \cite{Mazza2007},
\begin{equation}
 \frac{D_r}{D_t}=\frac{3}{4R^2}, \label{eq.coeff_rel}
\end{equation}
and so the rotational damping time $\tau_r=10\tau_t/3$.

The parameters for the Langevin dynamics simulations were taken from
Ref. \cite{Wang2012}: $m=10^{-12}g$ and $R=0.5\mathrm{\mu m}$. To access time scales of the order of the second we used
translational and rotational diffusion coefficients $D_t\approx6\mathrm{\mu m^2/s}$ and 
$D_r\approx18s^{-1}$, which are one order of magnitude larger than the ones observed experimentally in solution at room temperature \cite{Wang2012}.
There are, at least, two relevant time scales governing the relaxation process: The typical time for a bond to 
break and the diffusion time. During our simulations we did not observe any bond breaking event, as $k_BT/\epsilon$ is low. 
By contrast, the relaxation dynamics consists of particle rearrangements leading to the formation of new bonds. 
Thus, in this low temperature limit, considering a larger diffusion coefficient corresponds to re-scaling the time.

The initial structures were generated using the stochastic model first presented in Ref.~\cite{Dias2013}
where bonds are considered irreversible during the growth and the mechanism of mass transport to the substrate
is advective \cite{Dias2013b}. We measure the total number of particles $N$ in
monolayers (ML) corresponding to $L^2$ particles, where $L$ is the lateral size of the substrate in 
units of the particle diameter $2R$.
  
\section{Results}\label{sec.res}

  \begin{figure}[t]
   \begin{center}
    \includegraphics[width=\columnwidth]{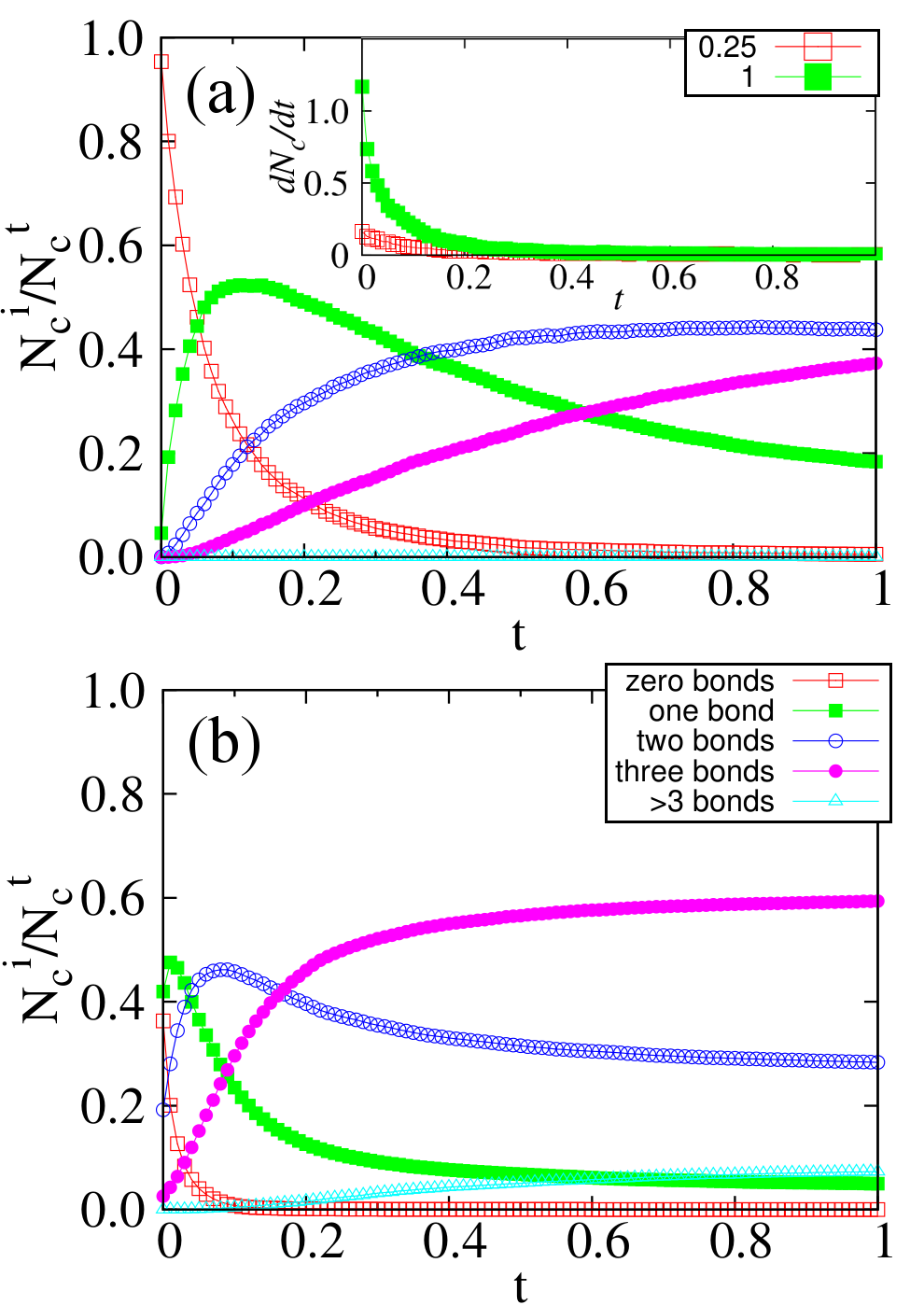} \\
\caption{Fraction of particles with a number of bonds $i$, for $i=\{0, 1, 2, 3, >3\}$, as a function of time for
(a) $N=0.25$ and (b) $N=1$ on a substrate of lateral size L=64, averaged over 10 samples. Inset: The binding rate,
 given by the derivative of the total number of bonds ($\times10^5$), as a function of time for $N=0.25$ and $N=1$ on a substrate 
 of size L=64.}
  \label{fig.numb_connections_substrate}
   \end{center}
  \end{figure}

   \begin{figure}[t]
   \begin{center}
    \includegraphics[width=\columnwidth]{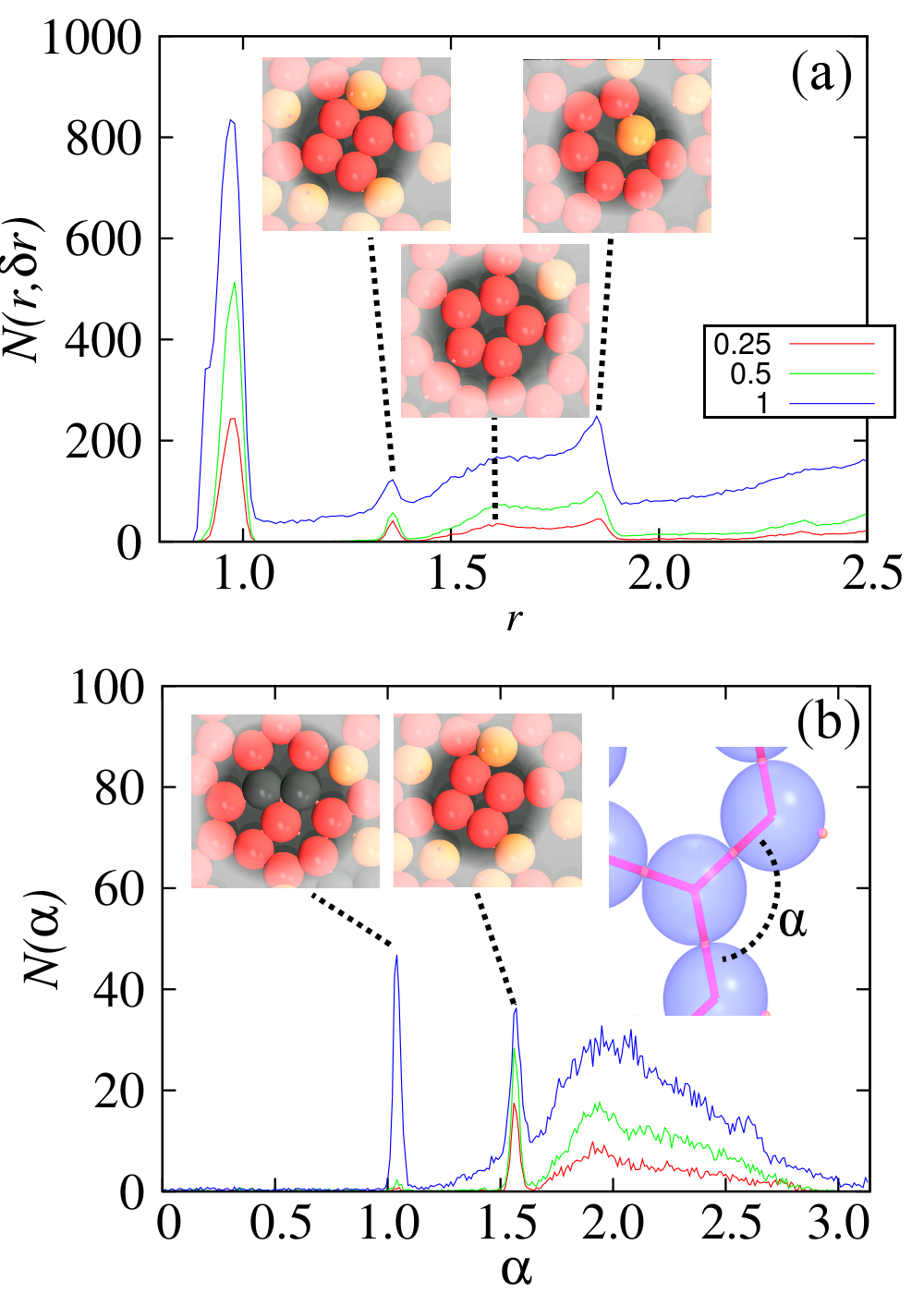} \\
\caption{(a) Radial distribution function $N(r,\delta r)$, where $r$ is the particle distance and
$\delta r=0.001$ is the size of the bins, for $N=\{0.25,0.5,1\}$
on a substrate of lateral size L=64, averaged over 10 samples. (b) Distribution function of
the angle $\alpha$ between three connected particles for $N=\{0.25,0.5,1\}$
on a substrate of lateral size L=64, averaged over 10 samples.}
  \label{fig.radial_distribution}
   \end{center}
  \end{figure}
  
 \begin{figure*}[t]
   \begin{center}
    \includegraphics[width=1.8\columnwidth]{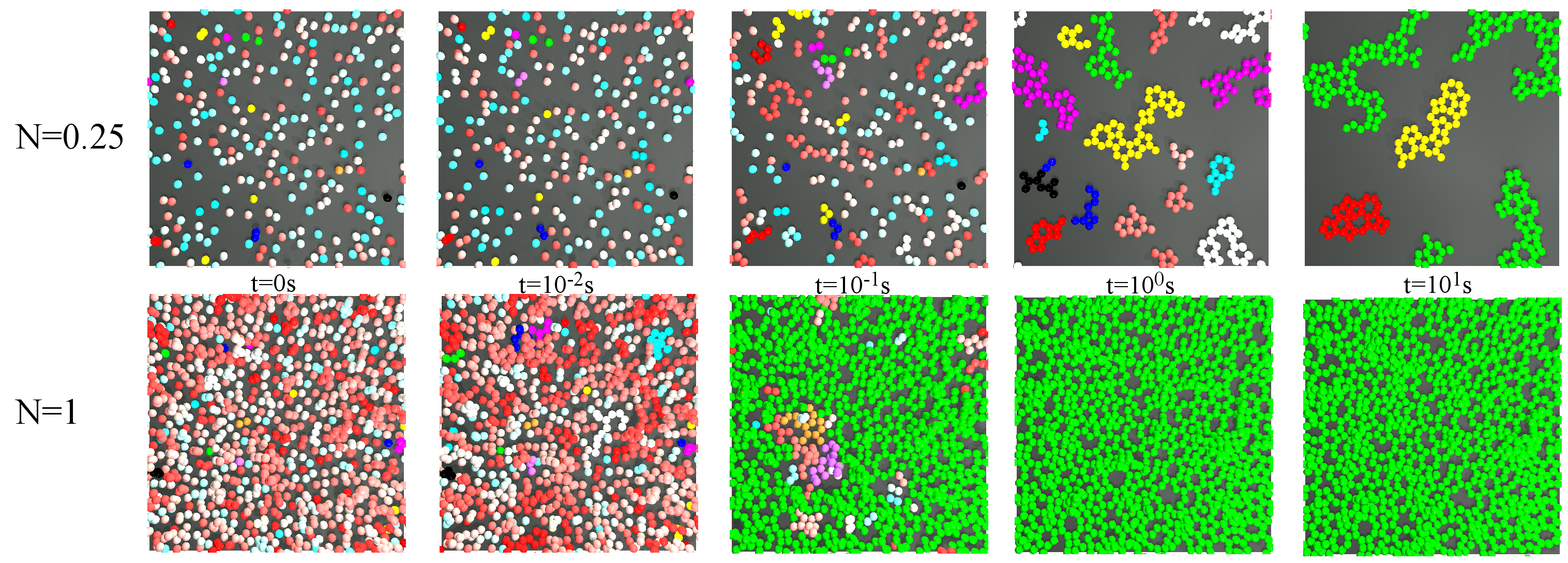} \\
\caption{Snapshots for $N=0.25$ and $N=1$. 
Images generated at $t=\{0, 10^{-2}, 10^{-1}, 10^0, 10^1\}$ seconds
on a substrate of lateral size L=32. Particles of the 
same color belong to the same aggregate of connected particles.}
  \label{fig.substrate_snaps_cluster}
   \end{center}
  \end{figure*}

\subsection{Particle bonds}

We analyze the dynamics for $N=0.25$ and $N=1$. From the 
snapshots in Fig.~\ref{fig.substrate_snaps_con} one can see that for the lower $N$ the overall structure 
evolves from an initially homogeneous distribution of patchy particles to a handful of aggregates. We define
aggregates as particles connected through their patches.
For $N=1$ the spatial distribution of particles does not change significantly in time and only a 
single aggregate spanning the entire substrate is formed. This result hints at two different mechanisms of relaxation. 
While for lower $N$, the particles need to diffuse and rotate to maximize the coordination of each 
particle, and thus decrease the overall energy, for larger $N$, the diffusion of the particles is hindered 
by the particle-particle repulsive interaction and relaxation is mainly driven by the local rearrangements of 
individual particles.

In Fig.~\ref{fig.numb_connections_substrate} we plot the ratio of particles
of a certain number of bonds as a function of time. A qualitative different behavior is observed 
for the two values of $N$. The inset of Fig.~\ref{fig.numb_connections_substrate}(a) shows the binding rate, given 
by the time derivative of the total number of bonds. The initial binding rate for $N=1$ is one order of magnitude larger than for $N=0.25$, as the 
initial interparticle distance on the substrate is lower and particles can promptly form bonds with their neighbors. This is also visible from the evolution of 
the fraction of isolated particles (unbonded) in the main plots of Figs.~\ref{fig.numb_connections_substrate}(a)~and~(b). For $N=1$ (Fig.~\ref{fig.numb_connections_substrate}(b)), this fraction vanishes for $t\approx0.1$, 
while for $N=0.25$ Fig.~\ref{fig.numb_connections_substrate}(a)) it takes six times longer for every 
isolated particle to form at least one bond. The fraction of particles with one and two bonds is characterized by a 
maximum at an intermediate time. As the initial relaxation dynamics is faster for $N=1$ than $N=0.25$, these maxima occur much earlier. 
For longer times, the fraction 
of particles with three bonds dominates (at low temperatures and strong patch-patch interaction). 
The fraction of particles with three bonds is higher for $N=1$, as only one aggregate is formed. 
Note that, for N=1, about $10\%$ of particles have a final number of bonds larger than three (the number of patches), while there are almost none for $N=0.25$. 
This result suggests that for a sufficiently large number of particles on the substrate the interparticle repulsion is such that multiple bonds per patch are formed to decrease the overall energy.

One expects that three-equally-spaced-patch particles self-organize into a honeycomb-like structure
\cite{Dias2015}. However, the radial distribution function, shown in
Fig.~\ref{fig.radial_distribution}(a), consists of a sequence of peaks at positions that differ from the 
characteristic interparticle distances of the honeycomb lattice.
The first peak of the radial distribution function is given by particles in direct contact. The second one 
indicates a square-like arrangement of connected particles, 
however they are not periodically reproduced, since the third and fourth neighbors of the 
square lattice are not present on the radial distribution function. The following peaks of the radial
distribution function range from a five-particle loop to more stretched ones of six, seven, or eight
particles. 

Figure~\ref{fig.radial_distribution}(b) shows the distribution of angles between two neighbors connected to
one particle (see the inset scheme). The 
first peak indicates the occurrence of double bonds on a single patch (see inset snapshot). This peak occurs at 
$\alpha=\pi/3$ suggesting that the local structure resembles a triangle. This is an uncommon
event since the probability that one particle has, at least, one double bond is $10\%$ (see discussion above). 
This event has a large effect on the angular distribution as it contributes three times to the statistics, one for each of the 
three particles. These structures are not visible in the radial distribution function 
of Fig.~\ref{fig.radial_distribution}(a) since the distance between the particles is one diameter. The next 
peak on the angular distribution function is related to square structures ($\pi/2$), and the widest peak
is the superposition of the contribution of the other structures forming loops of five to eight particles. This distribution
is considerably different from the original structure from the stochastic model (without relaxation), which exhibits
only one peak at $\alpha=2\pi/3$ \cite{Dias2013}.

\subsection{Aggregation dynamics}\label{sec.cluster_agg}

We proceed to analyze the aggregation dynamics. 
We label particles belonging to the same 
aggregate using the Hoshen-Kopelman algorithm \cite{Hoshen1976}. As shown in 
Fig.~\ref{fig.substrate_snaps_cluster} for $N=1$ the dynamics evolves to
form a single aggregate. For $N=0.25$ the formation of a single 
aggregate may occur at long enough times, if the patch-patch interaction
is strong enough.

  \begin{figure}[t]
   \begin{center}
    \includegraphics[width=\columnwidth]{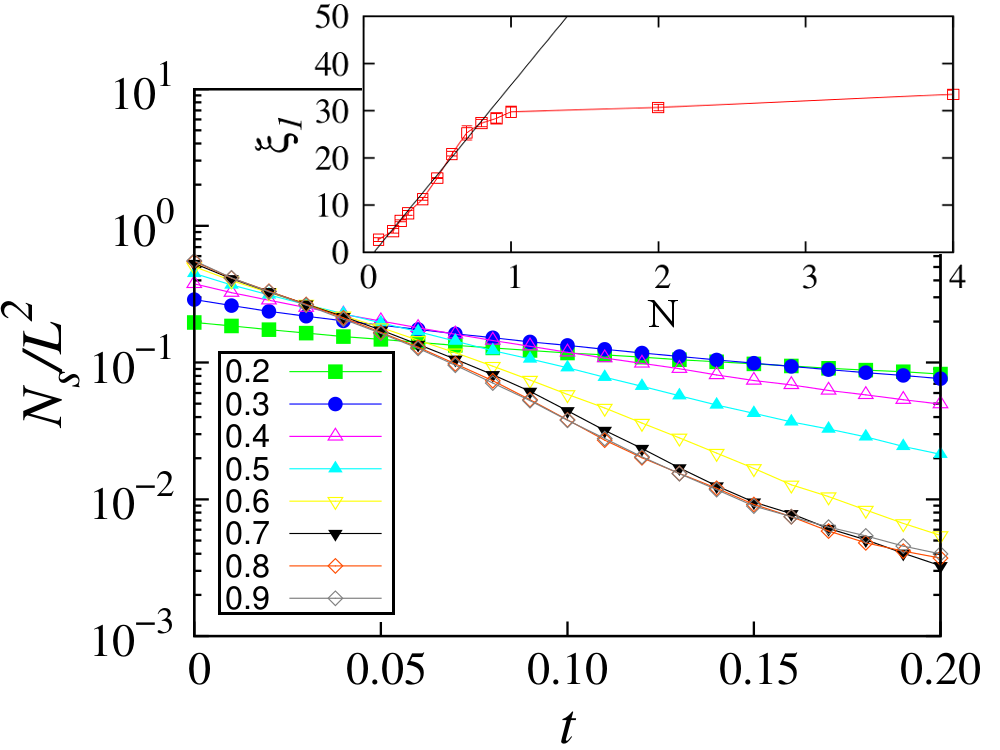} \\
\caption{Exponential decay, at initial times, for the number of aggregates 
as a function of time for $N$ from $N=0.2$ to $N=0.9$, on a 
substrate of lateral size L=64 and averaged over 10 samples. Inset: Exponential decay
$\xi_1$ as a function of $N$.}
  \label{fig.xi1_first}
   \end{center}
  \end{figure}

\begin{figure}[t]
   \begin{center}
    \includegraphics[width=\columnwidth]{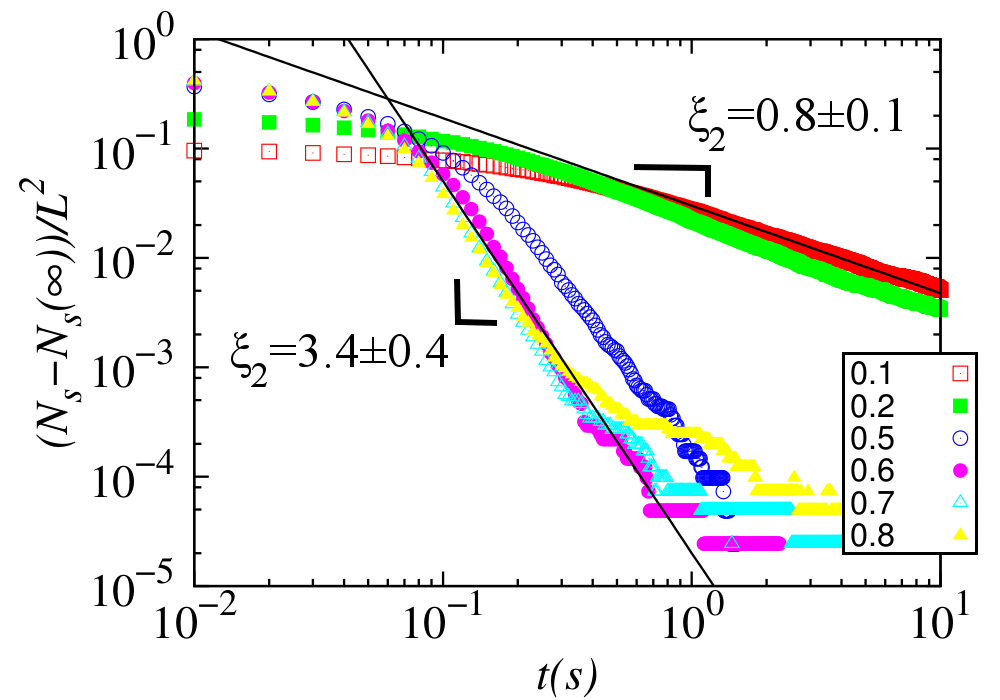} \\
\caption{Number of aggregates as a function of time for 
$N=\{0.1, 0.2, 0.5, 0.6, 0.7\}$, on a substrate of lateral size L=64 
and averaged over 10 samples.}
  \label{fig.xi2_second}
   \end{center}
  \end{figure}

We measure the number of aggregates $N_s$ as a function of time for different $N$, as shown in Figs.~\ref{fig.xi1_first}~and~\ref{fig.xi2_second}. 
For a strong patch-patch interaction and very long times (and below the bond breaking time), all particles are connected, forming
a single aggregate. From these results, we can distinguish two regimes: the first regime for $t<0.2s$ and the second for $t>=0.2s$.
In Fig.~\ref{fig.xi1_first}, we plot the number of aggregates as a function
of time for the first regime to show that $N_s$ has an exponential
decay,
\begin{equation}
 N_{s1}(t)\sim \exp(-\xi_1 t), \label{eq.exp_decay} 
\end{equation}
where $\xi_1$ is the inverse of a characteristic time scale.
As we can see from the snapshots in Fig.~\ref{fig.substrate_snaps_cluster},
in the first regime, the dynamics is dominated by fast formation of bonds with local neighbors leading to the
formation of small aggregates.

In the inset of Fig.~\ref{fig.xi1_first} we plot $\xi_1$
as a function of $N$. For low $N$, $\xi_1$ increases
linearly with $N$.
In the submonolayer regime, particles are initially distributed on the substrate in a homogeneous fashion. They diffuse and form aggregates. 
The typical area that each isolated particle needs to diffuse and aggregate scales with the inverse of the total number of particles on the substrate ($N$). 
Since $\xi_1$ is the inverse of the characteristic time it should scale linearly with $N$. 
For $N=0.8$ (and above) as increasing the number of particles does not reduce the typical
inter-particle distance in the first-layer on the substrate due to the nucleation of the second layer, $\xi_1$ saturates and remains practically 
independent of $N$.

Figure.~\ref{fig.xi2_second} shows the 
approach to the asymptotic value $N_s(\infty)=1$ for different values of $N$. 
We can see a power-law behavior at long times,
\begin{equation}
 N_{s2}(t)-N_s(\infty)\sim t^{-\xi_2}, \label{eq.pow_law_decay} 
\end{equation}
which reveals the second relaxation regime. As can be seen in the snapshots
of Fig.~\ref{fig.substrate_snaps_cluster}, the overall structure does not change significantly
in this regime. From Fig.~\ref{fig.xi2_second} we observe two different power laws 
for small and large $N$. For $N=0.1$ and $N=0.2$, an exponent of $\xi_2=0.8\pm0.1$ is found, indicating 
very slow aggregation dynamics. For $N>0.4$, we found an exponent of $\xi_2=3.4\pm0.4$, indicating
a faster relaxation. For values of $N=0.3$ and $N=0.4$, crossover effects are observed suggesting 
that the transition between the two regimes occurs around these values.
  
\begin{figure}[t]
   \begin{center}
    \includegraphics[width=\columnwidth]{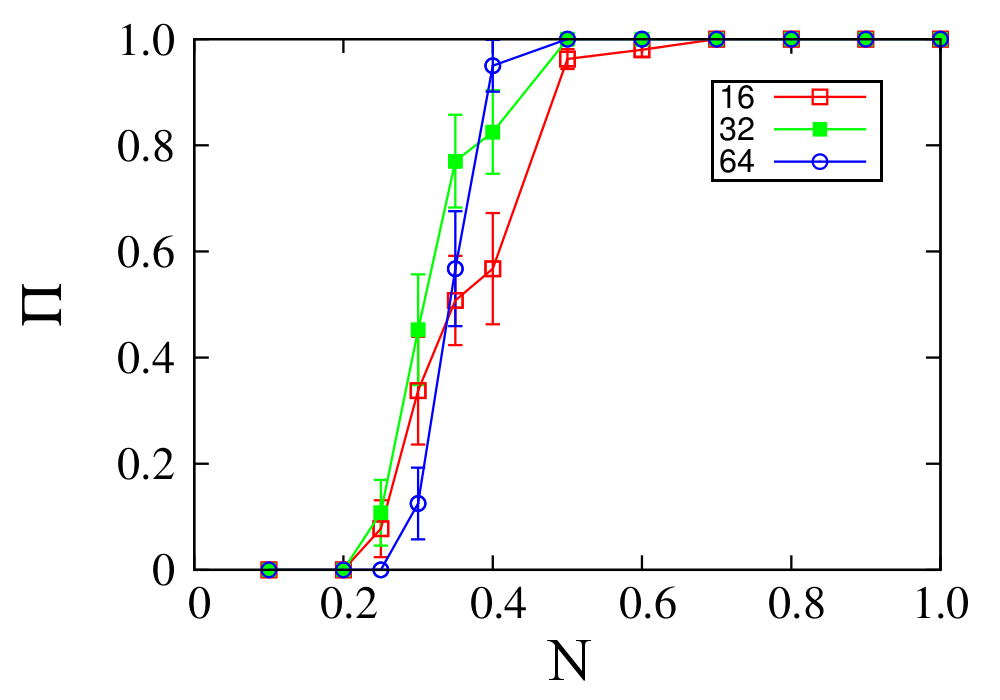} \\
\caption{Wrapping probability $\Pi$ as a function of $N$. 
Results are averages over 20 samples for substrates of lateral size of $L=\{16, 32, 64\}$.}
  \label{fig.wrapping}
   \end{center}
  \end{figure}

The power-law scaling has one of two characteristic exponents depending on $N$. 
For small $N$, aggregates need to diffuse and eventually merge. While for larger $N$, 
diffusion is negligible and aggregates need to readjust to merge with neighboring ones. 
This transition can be related to a percolation transition. 
In Fig.~\ref{fig.wrapping}, we plot the wrapping probability, given by the probability of finding an aggregate 
that spans the system from one side to the other (the same aggregate crosses opposite boundaries along the x-direction) 
as a function of $N$, for three different sizes. The percolation transition only
occurs for $N>0.3$ which is in agreement
with the transition found in the second relaxation regime shown in Fig.~\ref{fig.xi2_second}
but below the mean-field solution of $0.5$ for a Bethe lattice with three neighbors \cite{Stauffer1994}.

\begin{figure}[t]
   \begin{center}
    \includegraphics[width=\columnwidth]{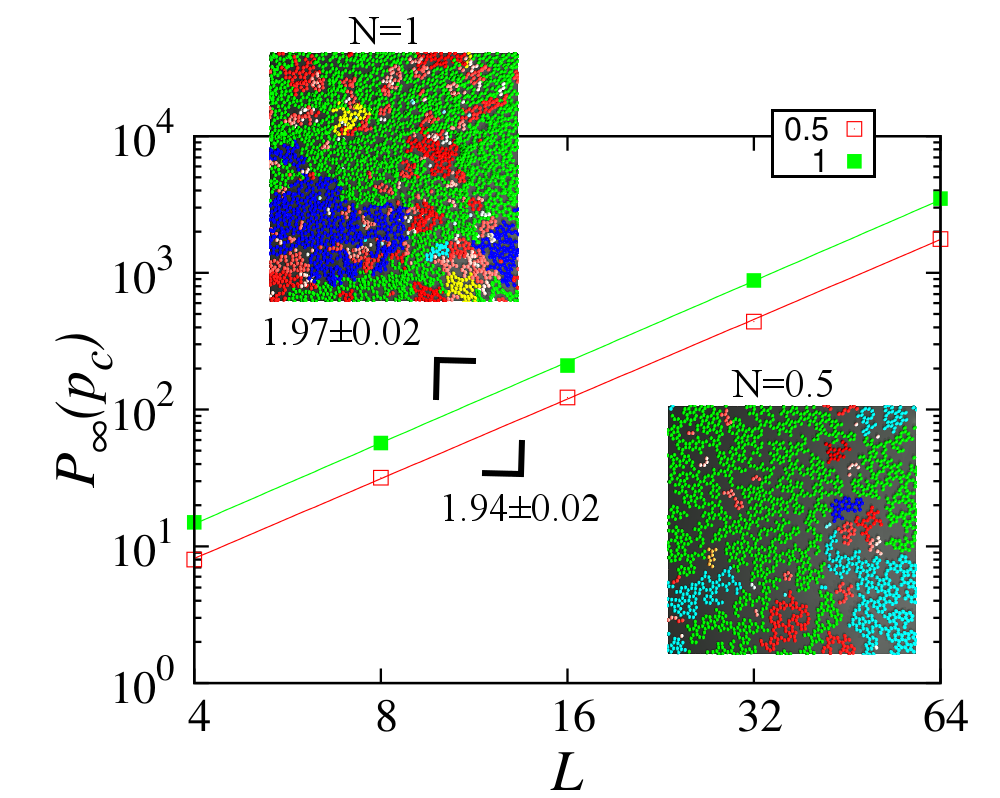} \\
\caption{Fraction of particles in the spanning aggregate ($P_\infty$) at the percolation threshold 
as a function of the substrate size for $N=0.5$ (open) and $N=1$ (filled). 
Results are averages over 10 samples of substrate size ranging from L=4 to L=64.}
  \label{fig.perc_cluster_size}
   \end{center}
  \end{figure}

The percolation threshold was estimated for
$N=0.5$ and $N=1$ by the peak on the order parameter standard deviation. From the position
of the peak we can estimate the percolation threshold in the thermodynamic limit by extrapolating for an infinite substrate.
To characterize the spanning aggregate we plot in Fig.~\ref{fig.perc_cluster_size} its size at the percolation threshold 
as a function of the substrate lateral size L for $N=0.5$ and $N=1$. It
scales as,
\begin{equation}
 S_\infty\sim L^{d_f}, \label{eq.fractal_dim}
\end{equation}
where $d_f$ is the fractal dimension. For $N=0.5$ the fractal dimension is $d_f= 1.94\pm 0.02$, 
in agreement with that for two-dimensional random percolation.
The numerical value of the fractal dimension for $N=1$ is slightly larger than the one expected for 
two-dimensional percolation, due to the contribution of particles from the second layer that also belong to the spanning aggregate.

\subsection{First-layer coverage}

\begin{figure}[t]
   \begin{center}
    \includegraphics[width=\columnwidth]{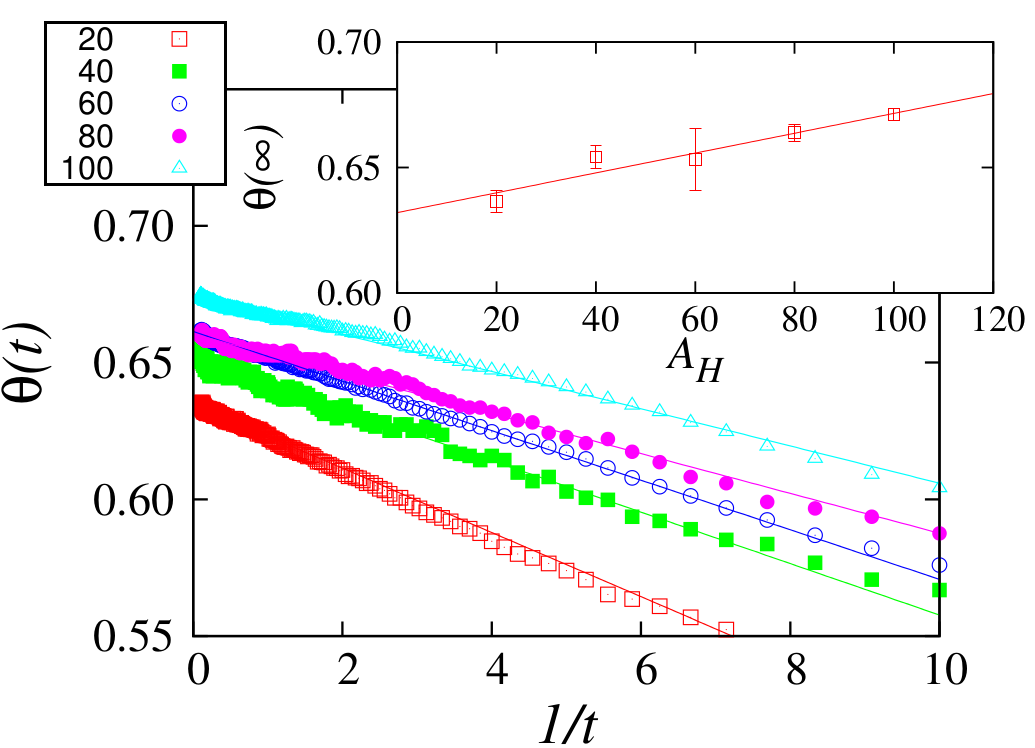} \\
\caption{Substrate first-layer coverage as a function of $1/t$ for $N=1$
and interaction strengths of the substrate 
potential of $A_H=\{20,40,60,80,100\}$ for a system of linear size L=32 and averaged over 10 samples. 
Inset: Asymptotic first-layer coverage as a function of the substrate potential interaction strength for 
$N=1$.}
  \label{fig.density_scaling}
   \end{center}
   \end{figure}
  
Figure~\ref{fig.density_scaling} shows the first-layer coverage for $N=1$ 
as a function of the inverse time. The first-layer coverage is defined as,
\begin{equation}
 \theta(t)=\frac{N_{subs}A_{col}}{L_xL_y}, \label{eq.packing_fraction}
\end{equation}
where $N_{subs}$ is the number of particles on the substrate within a distance of one diameter from it, 
$A_{col}$ the cross-section area of a sphere, and $L$ the 
lateral size of the substrate. $\theta$ is initially lower than the expected value for single-layer random sequential adsorption \cite{Evans1993a},
due to the formation of more than one layer \cite{Dias2013}. During the relaxation, $\theta$ increases, 
suggesting that particles adsorbed on top of others are attracted to the substrate.  
The coverage exceeds the one expected for a Honeycomb-like arrangement.
We extrapolate the first-layer coverage in the thermodynamic limit for different values of the strength of the particle-substrate interaction, 
$A_H$ (see inset of Fig.~\ref{fig.density_scaling}). We observe a 
dependence on $A_H$, as we increase the substrate interaction, the local restructuring 
increases the possibility of more particles being adsorbed on the substrate, increasing the first-layer coverage. The radial distribution 
function for various $A_H$ indicates some inter-particle penetration, a larger peak for the square structures, and a more enhanced peak related
to the contribution of the loops of five to eight particles (not shown here).
   
\begin{figure}[t]
   \begin{center}
    \includegraphics[width=\columnwidth]{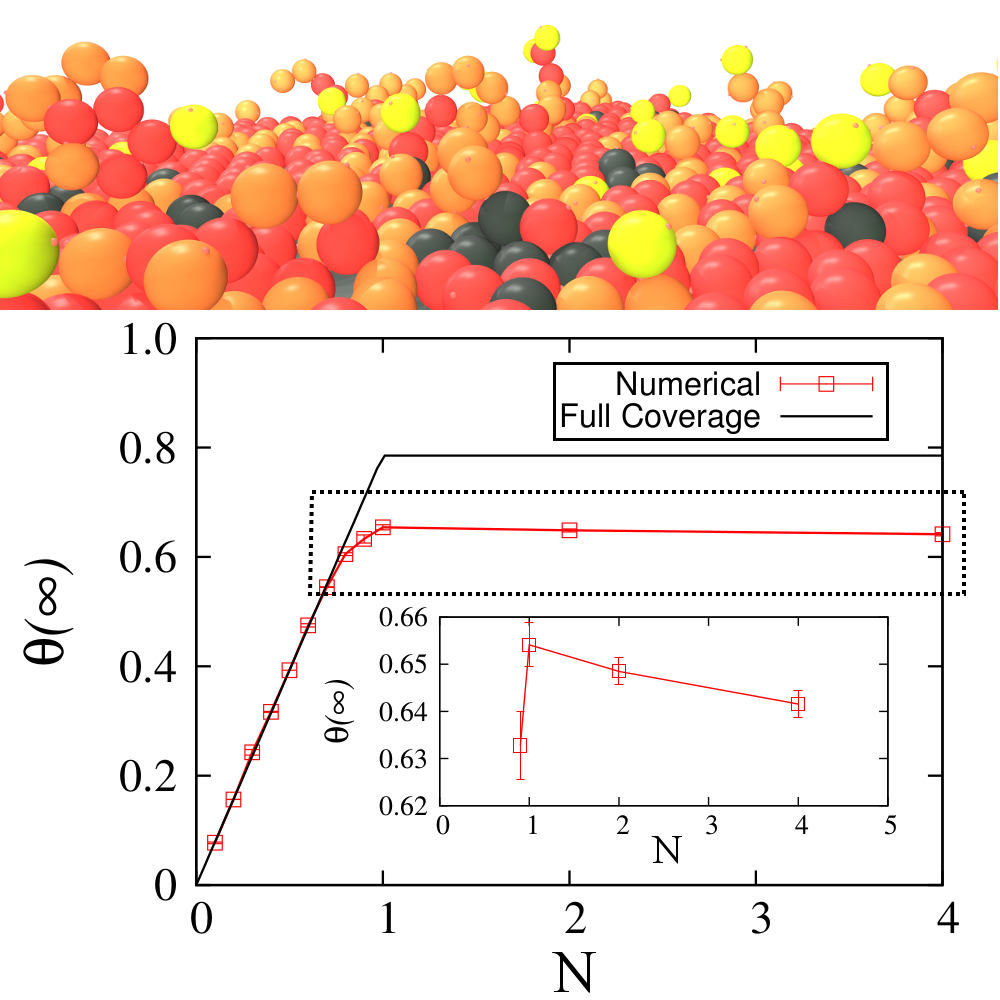} \\
\caption{Asymptotic first-layer coverage as a function of $N$
for a particle-substrate interaction strength
of $A_H=40$. Inset: Magnified region from $N=1$ to $N=4$. 
Top snapshot: Structure formed above 
de substrate layer that can hinders the access of more particles to the substrate.}
  \label{fig.density_layers_effect}
   \end{center}
  \end{figure}
   
In Fig.~\ref{fig.density_layers_effect} the extrapolated first-layer coverage as a 
function of $N$ is plotted. We observe the 
predictable increase in the submonolayer case, since all particles eventually move to the first layer. However, for larger number of particles, 
the first-layer coverage decreases with $N$ (see inset of Fig.~\ref{fig.density_layers_effect}). 
This is related to the formation of connected structures of
patchy particles that hold the particles away from the minimum of the particle-substrate potential (see snapshot 
in Fig.~\ref{fig.density_layers_effect}).

\section{Conclusions}\label{sec.conc}

Self-assembly of strongly interacting particles towards thermodynamic structures is typically hindered by the 
formation of kinetically trapped structures that are stable over long time scales. For the self-assembly of 
patchy colloids on an attractive substrate we considered the relaxation dynamics towards equilibrium, at low temperature, 
and identified the relevant kinetic structures. We have shown how the formation and 
relaxation of kinetic structures depends on the total number of particles and on the particle-substrate interaction. We observed a dependence
of the relaxation dynamics on the number of absorbed particles at both short and long times. We have found that the long time dynamics exhibits
two characteristic power-law exponents depending on the coverage.

Combining the directionality of particle-particle interactions and the presence of a substrate provides a 
promising route to grow (mono- and multi-layer) regular structures. 
However, our results reveal that, while strong patch-patch interactions favor the resilience to thermal 
fluctuations they yield large barriers to the relaxation towards equilibrium. Thus, as a follow up it would be of interest 
to explore possible annealing strategies (such as temperature annealing cycles) to overcome these barriers. 
Nevertheless, it is interesting to notice that these non-equilibrium structures are stable over long periods of time. 
A systematic analysis of their mechanical and optical properties might reveal them useful for certain practical applications.

\begin{acknowledgments}
We acknowledge financial support from the
Portuguese Foundation for Science and Technology (FCT) under Contracts
nos. EXCL/FIS-NAN/0083/2012, UID/FIS/00618/2013, and IF/00255/2013. 
\end{acknowledgments}

\bibliography{/home/cris/Copy/Research/Articles/Bibtex/SoftMatter}

\end{document}